\documentclass{elsart}

\usepackage{amssymb}

\begin{document}

\begin{frontmatter}

\title{Scaling in Small-World Resistor Networks}

\author[RPI]{G. Korniss\corauthref{cor1}},
\author[LANL]{M.B. Hastings},
\author[UH]{K.E. Bassler},
\author[UA]{M.J. Berryman},
\author[RPI]{B. Kozma},
\author[UA]{D. Abbott}

\corauth[cor1]{E-mail: korniss@rpi.edu}

\address[RPI]{Department of Physics, Applied
Physics, and Astronomy, Rensselaer Polytechnic Institute, 110
8$^{th}$ Street, Troy, NY 12180--3590, USA}
\address[LANL]{Center for Non-linear Studies and Theoretical Division,
Los Alamos National Laboratory, Los Alamos, NM 87545, USA}
\address[UH]{Department of Physics, 617 Science and Research Blvd I,
Univesity of Houston, Houston, TX 77204-5005, USA}
\address[UA]{Centre for Biomedical Engineering (CBME) and School of Electrical \&
Electronic Engineering, The University of Adelaide, SA 5005, Australia}

\begin{abstract}
We study the effective resistance of small-world resistor networks.
Utilizing recent analytic results for the propagator of the
Edwards-Wilkinson process on small-world networks, we obtain the
asymptotic behavior of the disorder-averaged two-point resistance
in the large system-size limit. 
We find that the small-world structure suppresses large network
resistances: both the average resistance and its standard deviation
approaches a finite value in the large system-size limit for any
non-zero density of random links. We also consider a scenario where
the link conductance decays as a power of
the length of the random links, $l^{-\alpha}$. In this case we find
that the average effective system resistance diverges for any non-zero
value of $\alpha$. 
\end{abstract}

\end{frontmatter}

Resistor networks have been widely studied since the 70's as models
for conductivity problems and classical transport in disordered
media \cite{Kirkpatrick71,Kirkpatrick73,DERRIDA82,Harris86}. 
Related studies on fuse networks have been investigated
on random percolating lattices with various applications to breakdown
processes in condensed matter and materials systems, ranging from
brittle fracture to dielectric breakdown \cite{Hansen91,Herrmann_rev,fuse1,DUXBURY95,fuse2}.
 
Recent research on complex networks
\cite{BarabREV,MendesREV,NEWMAN_SIAM} has turned to focus on 
dynamics on networks with applications to
synchronization in natural and artificial systems
\cite{Strogatz_review,BARAHONA02,LAI03,MOTTER05,GRIN05,KORNISS03}, and
transport phenomena \cite{BARA03,TORO04,barrat,LAI05}. 
Interesting recent studies have examined the tradeoffs between
redundancy and pleiotropy \cite{Berryman04}, and centralized versus decentralized design
\cite{Ashton05}, in complex networks. 
Finding the resistance between
any two points on a complex network is tractable and builds upon early
mesh-resistance techniques \cite{Aitchison}.
Estimating the strength of collaborative ties between nodes in
collaboration networks \cite{NEWMAN01} and quantifying the centrality of a node in
weighted networks can also be modeled by resistor networks \cite{NEWMAN04}.
While resistor networks have been employed to study and explore community structures in
social networks \cite{NEWMAN04,HUBER04,NEWMAN05}, they have not been
investigated as prototypical models for transport phenomena in complex
networks until very recently \cite{Lopez2005,Rieger05}.
The work by L\'opez et al. \cite{Lopez2005} revealed
that in scale-free (SF) networks  \cite{BarabREV,Barab_sci} anomalous transport
properties can emerge, displayed by the power-law tail of distribution
of the network conductance. 

Here we investigate the effective system resistance of small-world
(SW) networks \cite{WATTS98,WATTS99,NEWMAN}. Our results, in part, are based on recent
calculations \cite{KHK04,KHK05,KHK_SPIE_2005,ee} of the disordered averaged propagator of the
Edwards-Wilkinson (EW) \cite{EW} process extended to a SW network. The
EW process on a network can be thought of in terms of a
synchronization paradigm in a noisy environment. 
As a linear approximation, it also serves as the simplest
model for generic causally-constrained queuing networks \cite{TORO_VIRTUAL}, such as
manufacturing supply chains, e-commerce based services facilitated by
interconnected servers \cite{Dong2005}, and certain
distributed parallel schemes on computer networks \cite{KORNISS03,KIRK2004}. In
the context of the latter, it was shown \cite{KORNISS03} that  when
extending the original short-range connections to a SW-like network
(essentially, by adding a small density of random links on top of a
regular graph), the spread between completion times of tasks performed on different nodes
of a computer network remains {\em bounded}, rather than diverging
over time. Further, an infinitesimal extra ``cost'' is sufficient to achieve
this reduction. An important measure of efficiency is the
spread (or ``width'') of task-completion landscapes in such processing 
networks (larger spread corresponds to longer delays and poorer
efficiency). It is evident that this measure---the width of the EW
landscape on a network---is identical to the average resistance
(characterizing transport efficiency) of the same
network.
While this connection between the network propagator and the network
resistance \cite{Kirkpatrick73,WU2004,Cserti2000}, just like the one
between random walks and network resistance
\cite{Doyle,Lovasz,Redner}, is well known, it has not been exploited to
study transport efficiency of SW networks. Further, the connection
between the average spread of an EW steady-state landscape and the resistance of the same
network gives some insight in treating synchronization and transport efficiency
on the same footing. Namely, understanding the effects of the SW links in suppressing
the diverging long-wavelength modes of the network propagator, originally
present in regular lattices.


Our main result is that in SW networks, the average system resistance becomes finite
for an arbitrarily small density of random links, governed by the same
behavior of the network propagator which is responsible for suppressing
``rough'' synchronization landscapes \cite{KORNISS03,KHK04}.


{\it The Edwards-Wilkinson process on a network.---}%
The EW process in a synchronization context on a network, is given by the Langevin equation
\begin{equation}
\partial_{t} h_i = - \sum_{j=1}^{N} A_{ij}(h_i-h_j) +
\eta_{i}(t)\;, 
\label{EW_ntwk}
\end{equation}
where $h_{i}(t)$ is the general stochastic field variable on a node
(such as fluctuations in the task-completion landscape in certain
distributed parallel schemes on computer networks
\cite{KORNISS03,KHK04}) 
and $\eta_{i}(t)$ is a delta-correlated noise with zero mean and variance
$\langle\eta_{i}(t)\eta_{j}(t')\rangle$$=$$2\delta_{ij}\delta(t-t')$.
Here, $A_{ij}$$=$$A_{ji}$$>$$0$ is the effective coupling between the nodes
($A_{ii}$$\equiv$$0$). Defining the network Laplacian,
$\Gamma_{ij}=\delta_{ij}\sum_{l}A_{il}-A_{ij}$, we can rewrite Eq.~(\ref{EW_ntwk})
\begin{equation}
\partial_{t} h_i = - \sum_{j=1}^{N} \Gamma_{ij} h_j +
\eta_{i}(t)\;.
\label{EW_ntwk_gamma}
\end{equation}
For the steady-state equal-time two-point correlation function one finds
\begin{equation}
G_{ij} \equiv \langle(h_{i}-\bar{h})(h_{j}-\bar{h})\rangle =
\hat{\Gamma}^{-1}_{ij} =
\sum_{k=1}^{N-1} \frac{1}{\lambda_{k}}\psi_{ki}\psi_{kj} \;,
\label{corr_func}
\end{equation}
where $\bar{h}=(1/N)\sum_{i=1}^{N}h_i$ and
$\langle\ldots\rangle$ denotes an ensemble average over the noise
in Eq.~(\ref{EW_ntwk_gamma}). Here, $\hat{\Gamma}^{-1}$ denotes the inverse
of $\Gamma$ in the space orthogonal to the zero mode. Also, $\{\psi_{ki}\}_{i=1}^{N}$ and
$\lambda_{k}$, $k=0,1,\dots,N-1$, denote the $k$th normalized eigenvectors and the corresponding
eigenvalues, respectively. The $k=0$ index is reserved for the zero mode of the
Laplacian on any network: all components are identical of this
eigenvector and $\lambda_{0}=0$. 
The last form in Eq.~(\ref{corr_func}) (the spectral decomposition of
$\hat{\Gamma}^{-1}$) is useful for exact numerical
diagonalization purposes.
As one can see from Eq.~(\ref{corr_func}), $G$ is the inverse of the coupling matrix $\Gamma$ in
the space orthogonal to the zero mode of the Laplacian.
In particular, the average spread or width in the synchronization
landscape becomes
\begin{equation}
\langle w^2 \rangle = 
\left\langle\frac{1}{N}\sum_{i=1}^{N}(h_i-\bar{h})^2\right\rangle =
\frac{1}{N}\sum_{i=1}^{N} G_{ii} =
\frac{1}{N}\sum_{k=1}^{N-1} \frac{1}{\lambda_{k}}\;.
\label{w2_def}
\end{equation}
For large networked systems, the above observable is typically
self-averaging $\langle w^2 \rangle\simeq[\langle w^2 \rangle]$, where
$[\dots]$ denotes the average over the network disorder. Thus, if one
is able to calculate the disorder-averaged propagator $[G_{ij}]$, it provides the
scaling behavior of the average spread of the synchronization
landscape in the limit of $N$$\to$$\infty$. 


{\it The two-point resistance of a network.---}%
The stationary currents and voltages in any network of resistors are
governed by Kirchhoff's and Ohm's laws
\begin{equation}
\sum_{n}A_{mn}(V_m-V_n) = I_{m}\;,
\label{Kirchhoff}
\end{equation}
where $A_{mn}$ is the conductance of the link between node $m$ and
$n$, and $I_{m}$ is the {\em net} current flowing {\em into} the network at
node $m$. Note that $I_{m}$ is zero, unless node $m$ is connected to an
external terminal. Connecting the network to a ``battery'' with fixed
voltage drop $V$ through nodes $i$ and $j$ as the input and output
terminals, yields 
\begin{equation}
\sum_{n}\Gamma_{mn}V_n = I(\delta_{mi}-\delta_{mj}) \;,
\label{K2}
\end{equation}
where $\Gamma_{mn}$ is the {\em same} network Laplacian as introduced
earlier in the context of the EW process [Eq.~(\ref{EW_ntwk_gamma})],
associating the link conductance with the coupling matrix there;
$\delta_{ij}$ is the Kronecker delta. Here, $I$ is the
magnitude of the current entering and leaving the system at 
node $i$ and node $j$, respectively.
Solving for the voltages is well defined in the subspace orthogonal
to the zero-mode of the network Laplacian; the right-hand side vector of
Eq.~(\ref{K2}) is in that subspace. Hence, introducing the voltages measured
from the mean $\hat{V}_{m} = V_{m}-\bar{V}$, where $\bar{V}$$=$$(1/N)\sum_{m=1}^{N}V_m$, 
and employing  $\hat{\Gamma}^{-1}$ one has
\begin{equation}
\hat{V}_{m} = 
\sum_{n} \hat{\Gamma}^{-1}_{mn} I_n =
\sum_{n} \hat{\Gamma}^{-1}_{mn}I(\delta_{ni}-\delta_{nj}) = I(G_{mi} - G_{mj})\;,
\label{voltage_solution1}
\end{equation}
where $G$ is the same network propagator discussed in the previous
section in the context of the EW process on networks [Eq.~(\ref{corr_func})].
Applying the above equation to the voltage difference across node
$i$ and node $j$ to which the battery is attached, one finds
\begin{equation}
V =  V_i-V_j=\hat{V}_i-\hat{V}_j = I \left( G_{ii} + G_{jj} -2G_{ij}\right) \;.
\label{voltage_solution2}
\end{equation}
For the equivalent two-point resistance between node $i$ and
$j$ one finally obtains
\begin{equation}
R_{ij} \equiv \frac{V}{I} = G_{ii} + G_{jj} -2G_{ij} =
\sum_{k=1}^{N-1} \frac{1}{\lambda_{k}}(\psi_{ki}^2 + \psi_{kj}^2 - 2 \psi_{ki}\psi_{kj}) \;,
\label{R_solution}
\end{equation}
where the last form in Eq.~(\ref{R_solution}) is, again,  useful for exact
numerical diagonalization purposes. Looking at
Eq.~(\ref{R_solution}), one can realize that
the two-point resistance of a network between node $i$ and $j$ is the
same as the steady-state {\em height-difference} correlation function of the EW process
on the network
\begin{equation}
\langle (h_i-h_j)^2\rangle = G_{ii} + G_{jj} -2G_{ij} = R_{ij} \;.
\label{R_C}
\end{equation}
The height-difference correlation function is a standard observable in
surface-growth phenomena, extensively studied in the past two decades
\cite{BARA95},
so many of the answers for regular resistor networks can be obtained directly
by looking at the equivalent EW model on a $d$-dimensional substrate.
For example, in an infinite one-dimensional system, the resistance
between two nodes, separated by a distance $|i-j|$, diverges with the
separation as $R_{ij}=R(|i-j|)\simeq|i-j|$ \cite{WU2004,Cserti2000,BARA95}. 
Another trivial, but insightful,
relationship between the EW process and the resistor network can be
obtained by summing up Eq.~(\ref{R_C}) over all $i\neq j$ pairs,
yielding
\begin{equation}
\bar{R} \equiv\frac{1}{N(N-1)}\sum_{i\neq j}R_{ij} = 2\langle w^2 \rangle \;,
\label{R_w2}
\end{equation}
i.e., the average system resistance of a given network is twice the
steady-state width of the EW process on the same network.

{\it Effective Resistance of simple SW Networks.---}%
First, we consider ``simple'' SW resistor networks, where the conductance of
each link is identical, with unit value, for simplicity. 
When studying network-transport phenomena for
systems where physical links are subject to strong cost and
geometric constraints, this can be unrealistic and cost-prohibitive. 
For others, e.g., modeling
information flow in social networks
\cite{NEWMAN04,HUBER04,NEWMAN05,TORO04a}, 
this can be an acceptable starting
point, since ``long-range'' connections do not necessarily degrade
the information carrying capacity and the efficiency (e.g., influence)
of that link. 
We start with a one-dimensional ring with $N$ nodes (i.e., impose
periodic boundary conditions), and add a ``random'' link to each
pair of nodes, independently for each pair, with probability $p/N$. In
addition to the two nearest-neighbor connections, now each node, on average has $p$
random links, so $p$ is the density of random links. The resulting
network is essentially an Erd\H{o}s-R\'enyi (ER) network \cite{ER} on top of a
one-dimensional graph. This SW construction slightly differs from
the original Watts-Strogatz one \cite{WATTS98} where
random links are introduced through ``rewiring''. The resulting network,
however, has the same universal properties in the small-$p$, large-$N$
limit \cite{NEWMAN_WATTS,MONA}, and is also more amenable to
analytic approximations.

The coupling matrix for the {\em differences} of the
relevant variables [Eq.~(\ref{EW_ntwk}) and (\ref{Kirchhoff})] then becomes
\begin{equation}
A_{ij} = \delta_{i,j-1} + \delta_{i,j+1} + J_{ij}\;,
\label{A_sw_ntwk}
\end{equation}
where the matrix elements $J_{ij}$ are quenched random variables;
$J_{ij}$$=$$1$ with probability $p/N$ and $J_{ij}$$=$$0$ with
probability $(1-p/N)$. The corresponding Laplacian then can be written as 
\begin{equation}
\Gamma_{ij}=2\delta_{i,j} - \delta_{i,j-1} - \delta_{i,j+1} 
+ \delta_{ij}\sum_{l}J_{il}-J_{ij} \;.
\label{SWN_Laplacian}
\end{equation}
Equations~(\ref{A_sw_ntwk}) and (\ref{SWN_Laplacian}), with $J_{ij}$
defined above, correspond to identical (unit) conductance for each existing connection in the resistor
network. Our numerical scheme relied on the exact numerical diagonalization of
the SW network Laplacian $\Gamma$ in Eq.~(\ref{SWN_Laplacian})
\cite{numrec}. Our analytic results, asymptotically exact in the large
system-size limit, are straightforward applications of those of the EW
propagator on SW networks \cite{KHK04,KHK05,KHK_SPIE_2005,ee}.

\begin{figure}[t]
\vspace*{2.5truecm}
       \includegraphics{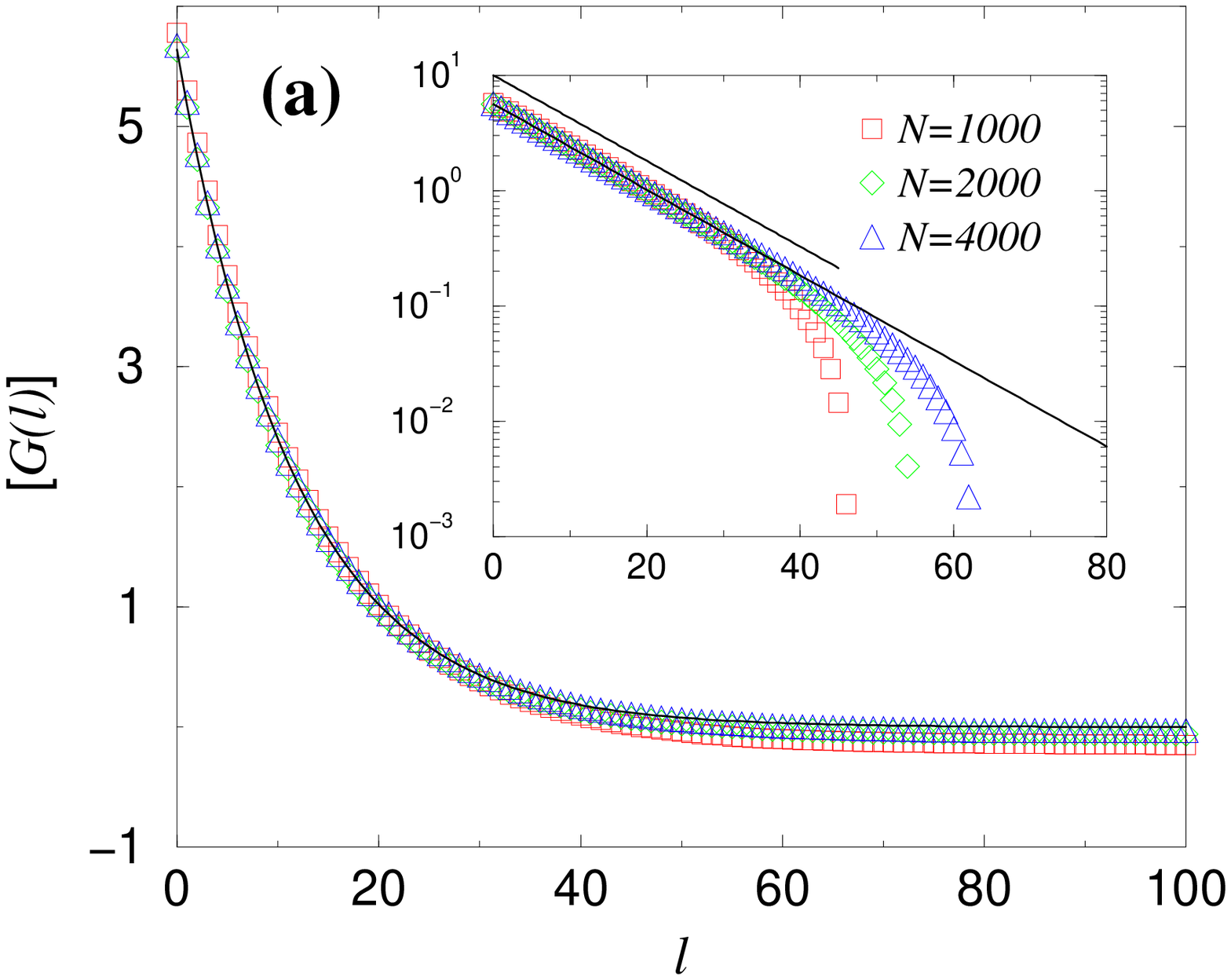}
       \includegraphics{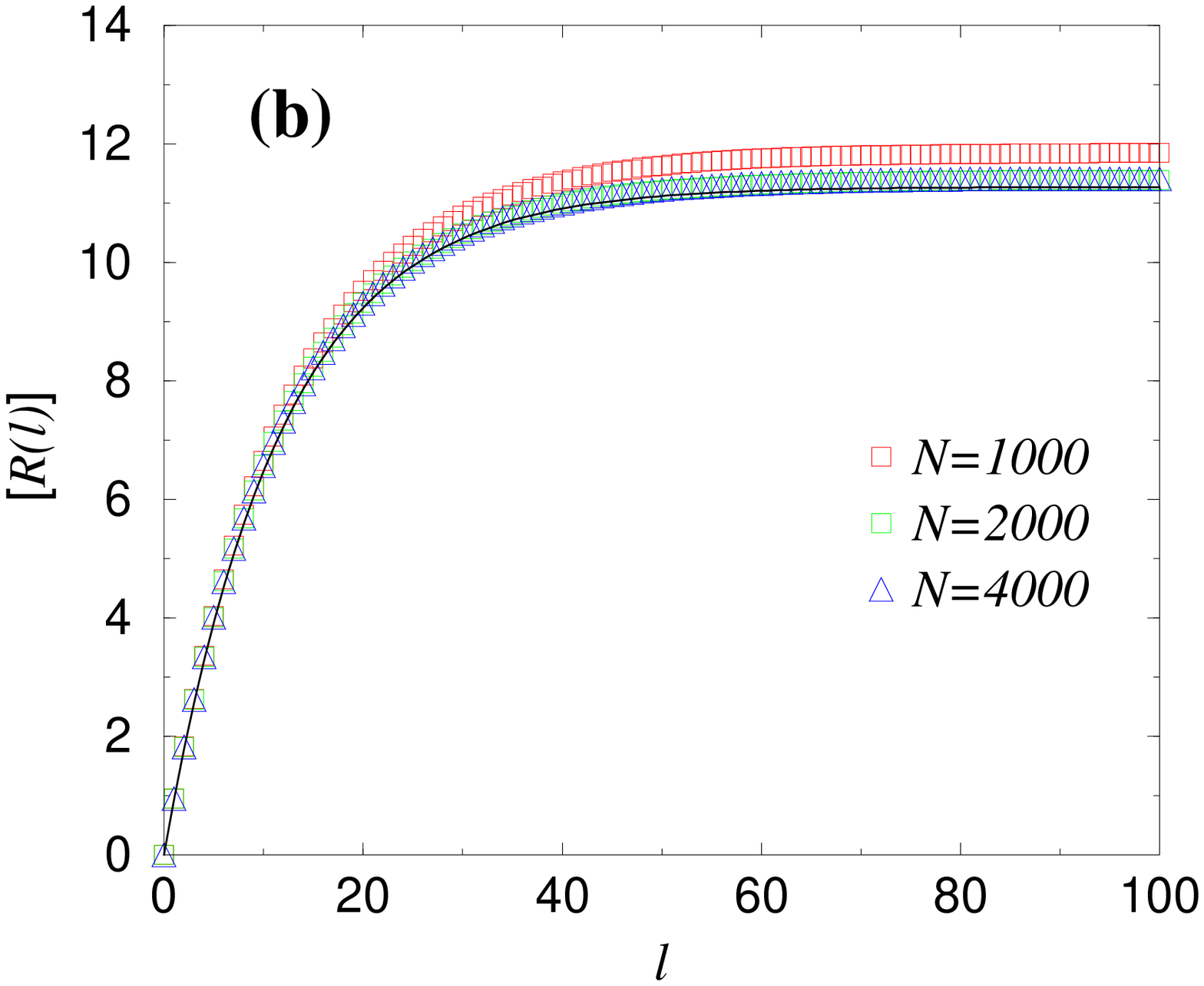}
\vspace*{3.20truecm}
\caption{Disorder-averaged (a) two-point function and (b) two-point resistance
as a function of the separation $l$ in simple SW networks for $p=0.10$
and three system sizes. The solid line in (a) and (b) corresponds to the exponential
decay and saturation given by Eqs.~(\ref{G_l}) and (\ref{R_l}), respectively}
\label{fig3}
\end{figure}

\begin{figure}[t]
\vspace*{2.5truecm}
       \includegraphics{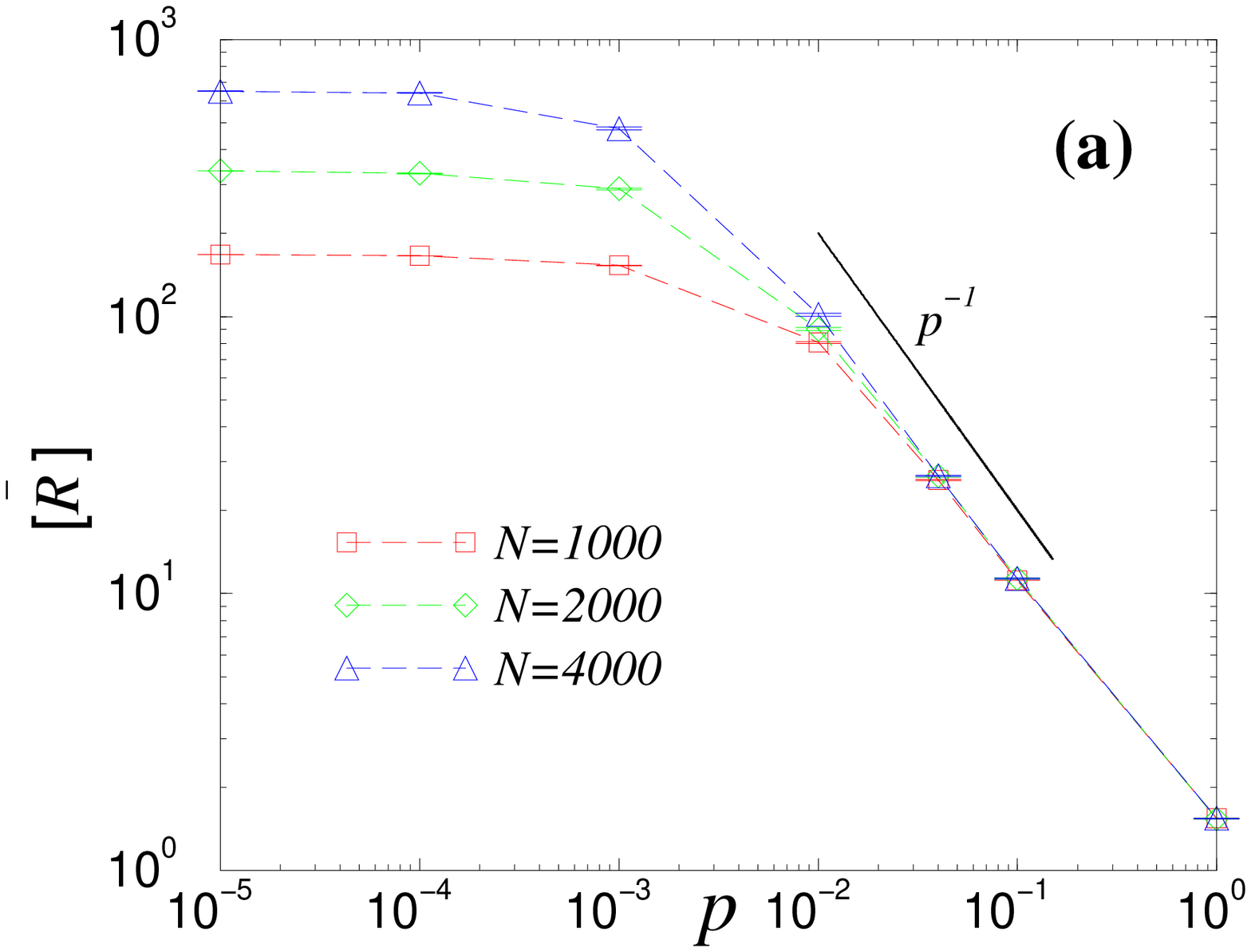}
       \includegraphics{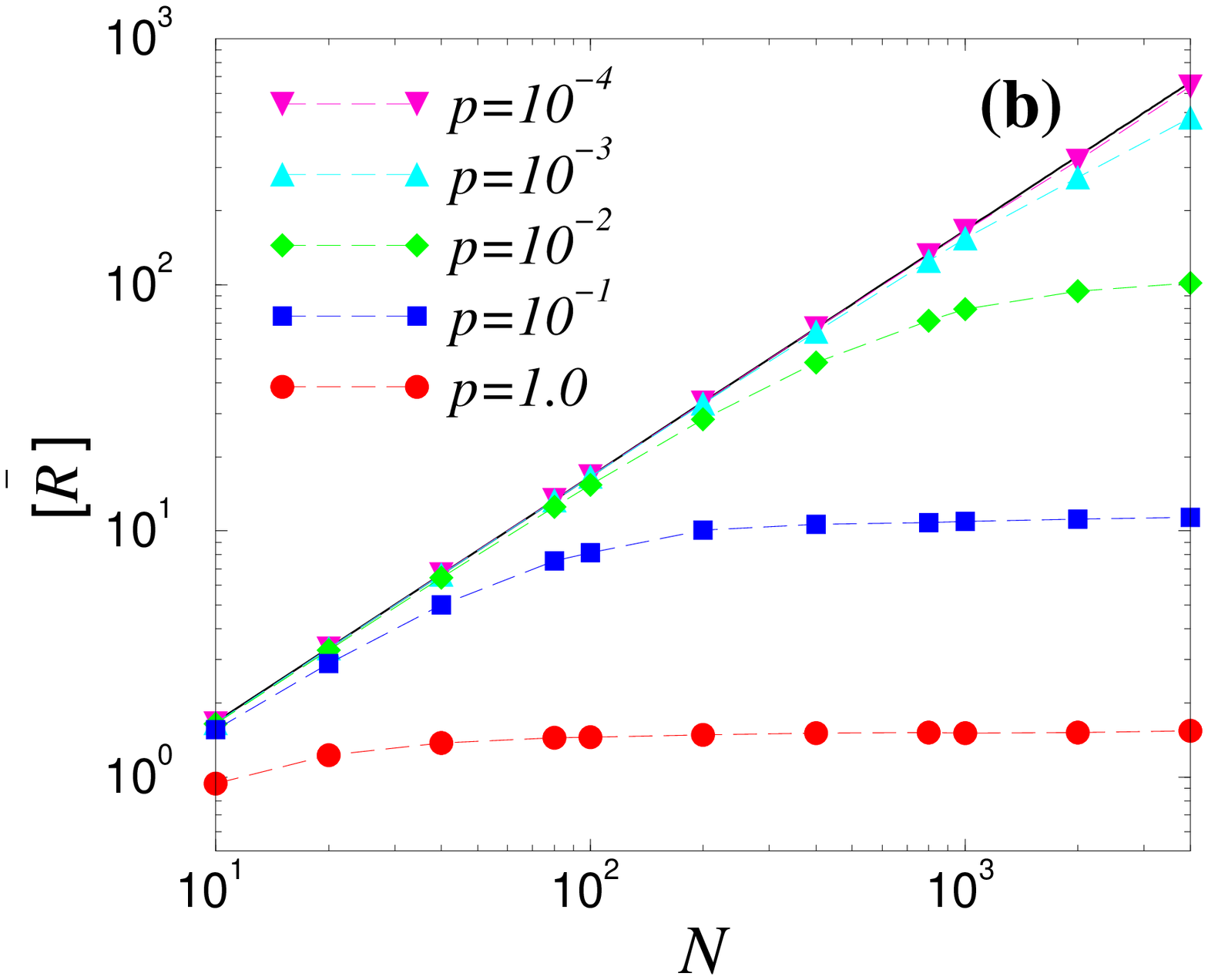}
\vspace*{3.20truecm}
\caption{(a) Average resistance vs the density of random links. The
       straight solid line indicate the asymptotic infinite
       system-size behavior [Eq.~(\ref{R_avg})].
(b) Average resistance vs the system size in simple SW networks. The
       straight solid line corresponds to the behavior of the
       one-dimensional regular network (ring), $[\hat{R}]\simeq N/6$.}
\label{fig1}
\end{figure}

Averaging over the network-disorder restores translational invariance, hence
the disorder-averaged two-point function $[G_{ij}]=[G(|i-j|)]$ will only
depend on the underlying Euclidean distance between the nodes.
These correlation functions have been calculated using
disorder-averaged self-consistent perturbation theory
\cite{KHK04,KHK05,KHK_SPIE_2005,ee}.
For the disorder-averaged two-point function for small $p$ values, in
the infinite system-size limit one finds \cite{KHK04,KHK05,KHK_SPIE_2005,ee} 
\begin{equation}
[G(l)] \simeq  \frac{1}{2\sqrt{\Sigma}}e^{-\sqrt{\Sigma} l} \;,
\label{G_l}
\end{equation}
where $\Sigma\sim p^2$ is an effective mass generated by the random
links for simple SW networks \cite{KHK04,MONA}. Then, using
Eq.~(\ref{R_solution}), for the average resistance on a SW network between
two nodes separated by a distance $l$, we obtain
\begin{equation}
[R(l)] = 2\left([G(0)] -[G(l)]\right) \simeq
\frac{1}{\sqrt{\Sigma}}\left(1-e^{-\sqrt{\Sigma} l}\right)
 \;,
\label{R_l}
\end{equation}
approaching a finite value in the limit of infinite separation,
$\lim_{l\to\infty}[R(l)]=\Sigma^{-1/2} \sim p^{-1}$.  
In contrast, on a regular one-dimensional ring, the resistance between
two nodes separated by a distance $l$ diverges in a power-law fashion,
$R(l) \simeq l$, as can be seen by taking the
$\Sigma$$\to$$0$ limit in Eq.~(\ref{R_l}) or by direct calculations on regular
lattices \cite{WU2004,Cserti2000}. 
Further, the average resistance is {\em finite}
for an arbitrarily small but non-zero $p$ in the limit of $N$$\to$$\infty$,
\begin{equation}
\bar{R}\simeq [\bar{R}] = 2[\langle w^{2} \rangle]= 2[G(0)] \simeq 
 \frac{1}{\sqrt{\Sigma}} \sim p^{-1}\;,
\label{R_avg}
\end{equation}
in strong contrast with average resistance for a regular network diverging as 
$\bar{R}\simeq N/6$. Equations~(\ref{R_l}) and (\ref{R_avg}) are the
central results of our paper. They capture the average resistance
between nodes separated by a distance $l$ and the average system
resistance for SW networks with a small but non-zero density of random
links, respectively.
Results from exact numerical diagonalizations, shown in
Fig.~\ref{fig3} and Fig.~\ref{fig1}, up to systematic finite-size effects,
agree very well with the above predictions.

\begin{figure}[t]
\vspace*{2.5truecm}
       \includegraphics{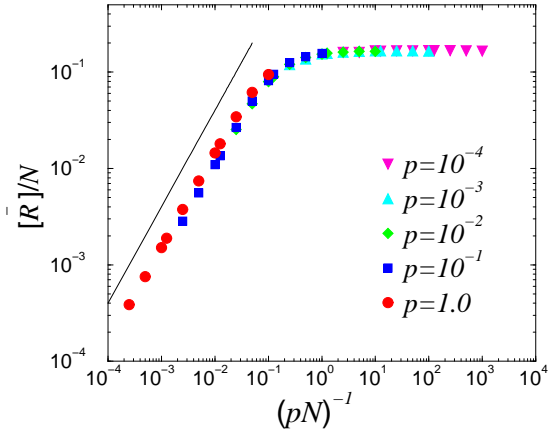}
\vspace*{3.40truecm}
\caption{Scaled average resistance in simple SW networks according to
Eq.~(\ref{R_fss}). The straight solid line
corresponds to the asymptotic behavior of the scaling function for
small arguments [Eq.~(\ref{f_fss})].}
\label{fig2}
\end{figure}

In addition to the above asymptotic results, valid in the infinite
system-size limit, we also constructed the scaling form,
capturing the finite-size effects, e.g., for the average resistance
\cite{KHK_SPIE_2005,KHK_PREP}.
From the above it is clear that in addition to
the linear system size $N$, there is one other lengthscale in the
problem for non-zero $p$ values, $\xi=1/\sqrt{\Sigma}\sim
p^{-1}$. This lengthscale is, in fact, the average distance between
nodes which have random links emanating from them.
For $p=0$ (the limit of a regular one-dimensional network) $[\bar{R}]\sim N$, while
for $p\neq 0$, in the infinite system-size limit, it approaches a constant,
$[\bar{R}] \simeq 1/\sqrt{\Sigma} =\xi \sim p^{-1}$ [Fig.~\ref{fig1}(a)
and (b)]. Thus, the finite-size behavior of the average resistance can
be expressed as
\begin{equation}
[\bar{R}] = N f(\xi/N)
 \;,
\label{R_fss}
\end{equation}
where $f(x)$ is a scaling function such that 
\begin{equation}
f(x) \sim \left\{ \begin{array}{ll}
x & \mbox{if $x\ll 1$} \\
{\rm const.} & \mbox{if $x\gg 1$} \;.
\end{array} \right.
\label{f_fss}
\end{equation}
The scaled numerical data, $[\bar{R}]/N$ vs $\xi/N$ [Fig.~\ref{fig2}],
shows good collapse, as suggested by Eq.~(\ref{R_fss}). 

\begin{figure}[t]
\vspace*{2.2truecm}
       \includegraphics{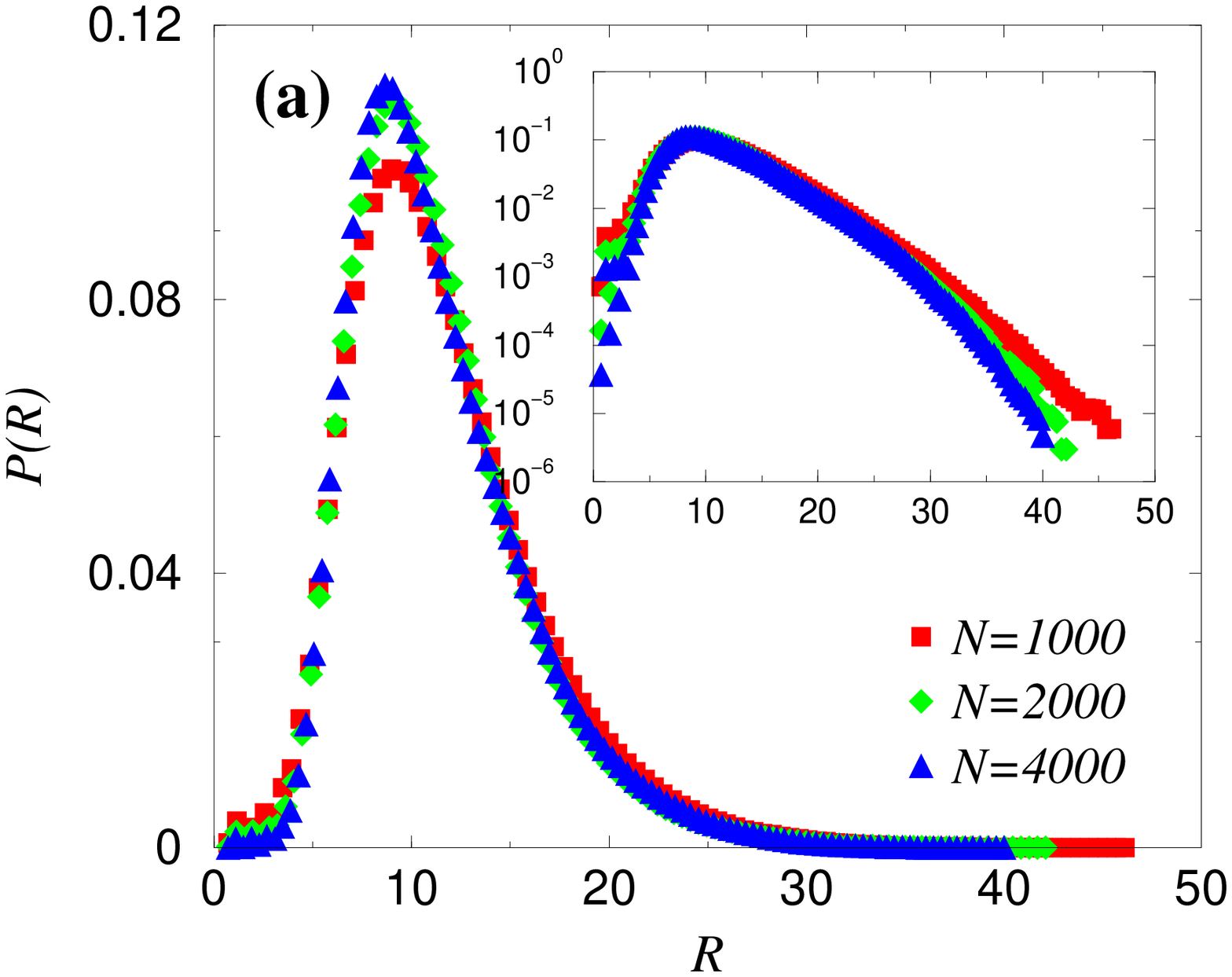}
       \includegraphics{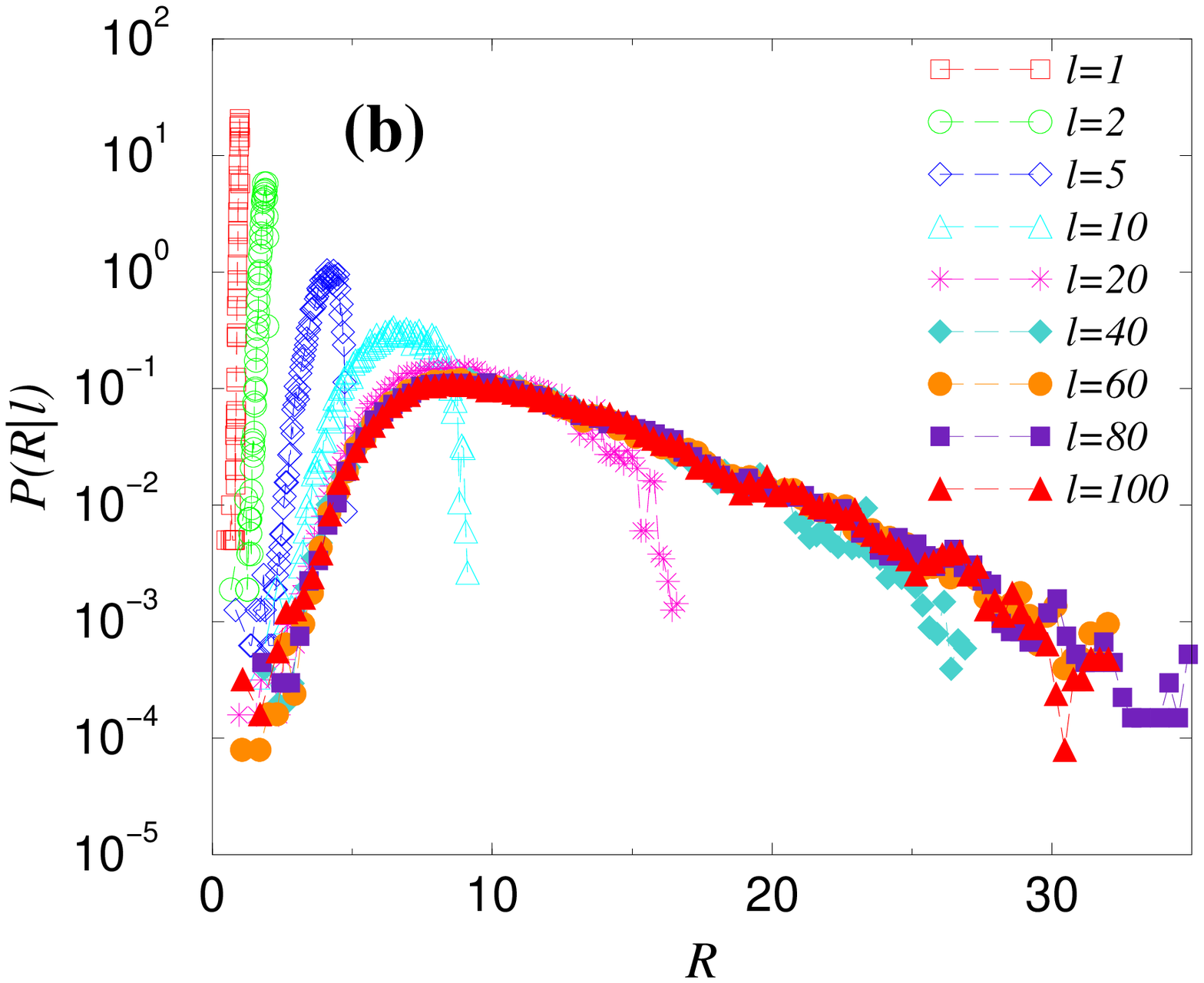}
\vspace*{5.7truecm}
       \includegraphics{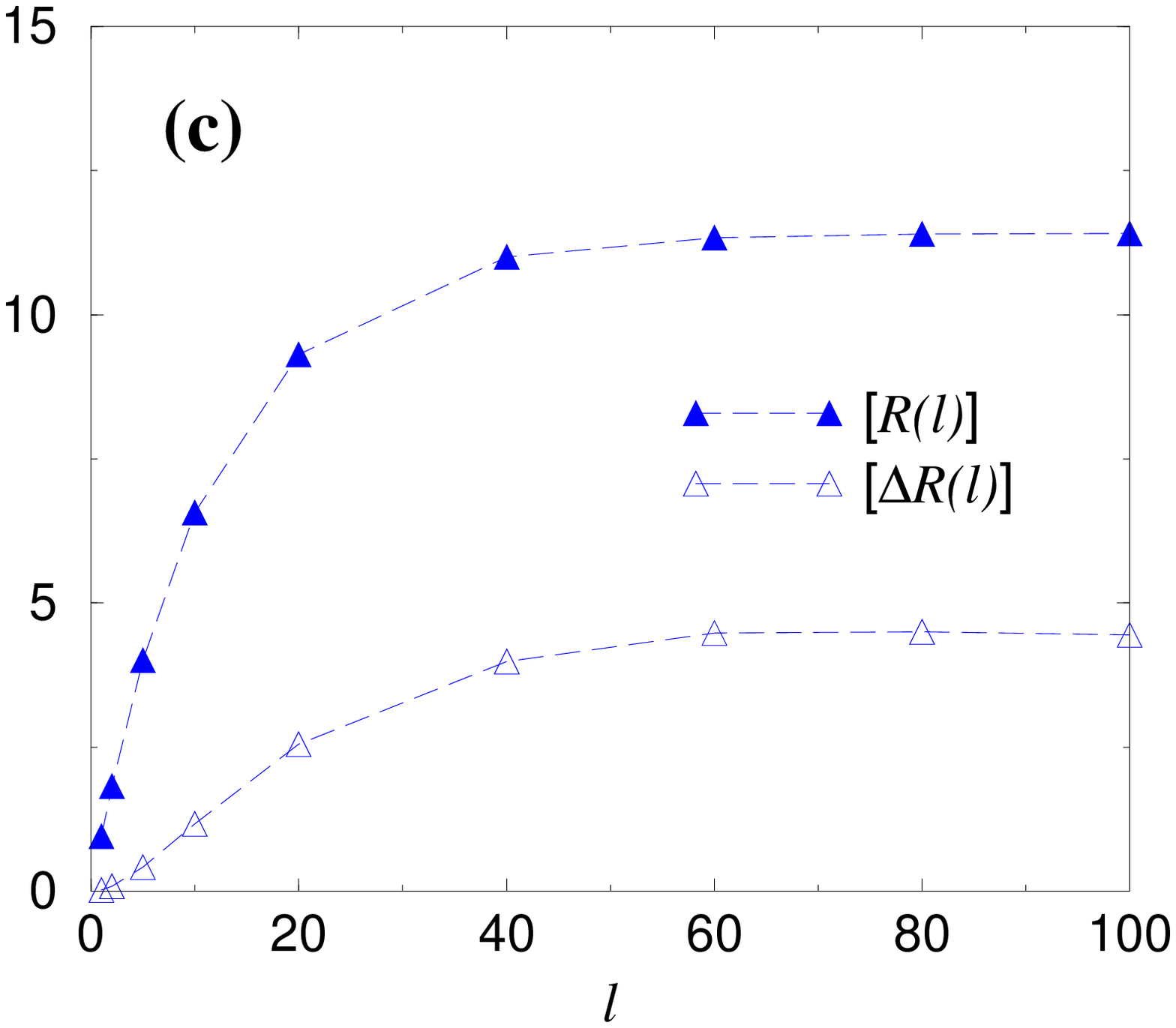}
\vspace*{3.20truecm}
\caption{(a) Network resistance probability distributions (histograms)
       for $p=0.10$ and for three system sizes. 
(b) Network resistance probability distributions (histograms) for nodes separated by a
distance $l$,  for various $l$ values for $N=4000$.
(c) Behavior of the average $[R(l)]$ and the standard deviation
$[\Delta R(l)]$ of the distributions shown in (b).}
\label{fig_P_R}
\end{figure}

\begin{figure}[t]
\vspace*{2.5truecm}
       \includegraphics{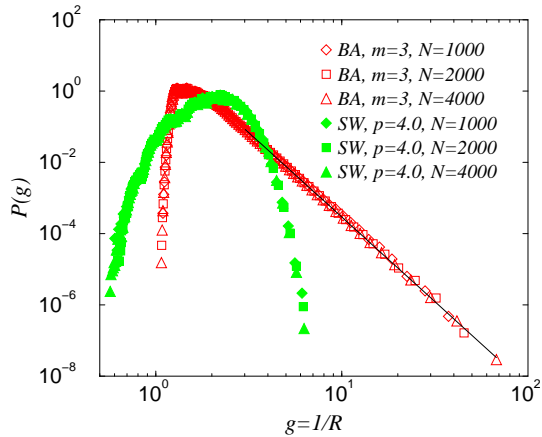}
\vspace*{3.40truecm}
\caption{Comparison of the network conductance distribution with that of
the BA scale-free network model with the same average degree $\langle
k\rangle$$=$$6$ (BA network with $m=3$ \cite{BA_m}, and SW network with  $p=4.0$.)}
\label{fig_BA}
\end{figure}

We also studied the probability distribution of the effective
resistance of the network [Fig.~\ref{fig_P_R}]. The overall
distribution is shown in Fig.~\ref{fig_P_R}(a).
Further, we constructed the distribution of the effective network resistance
between two nodes separated by a distance $l$, $P(R|l)$ \cite{Harris86}, 
indicating that they converge to a limit distribution for $l$$\to$$\infty$
[Fig.~\ref{fig_P_R}(b) and (c)].
These results imply that for SW networks, {\em
both} small and large effective resistance values are strongly (at
least exponentially) suppressed about the average. This is in strong
contrast with the behavior of SF resistor networks \cite{Lopez2005},  
where large resistance values are strongly suppressed, but the probability of
small values decays only in a power-law fashion; hence a power-law
tail in the conductance distribution occurs for large $g$$\equiv$$1/R$ values.
This finding for SF networks implies \cite{Lopez2005}, that there
exist a few nodes (``hubs'') in the system that, if selected as the input and output
nodes, can support anomalously large transport through the network.
This phenomenon is absent (as one can expect) in SW networks, just
like in completely random (ER) networks \cite{Lopez2005} related to
the exponential tail of the degree distributions  of these networks. In
Fig.~\ref{fig_BA} we compare the conductance distributions for the
SW and the Barab\'asi-Albert (BA) \cite{Barab_sci,BA_m} SF network, with the
same average degree and uniform link conductance.
 At this point we note that while anomalously large
conductances are absent in SW networks, they are more efficient, on average, in supporting
transport between two arbitrary pairs of nodes, i.e., $[R]_{SW} <
[R]_{SF}$, and  $[g]_{SW} > [g]_{SF}$. In comparison, for the two networks
shown in Fig.~\ref{fig_BA} with the same average degree and uniform
link conductance, the average
network resistance and conductance values are
$[R]_{SW}\simeq 0.472$,  $[R]_{SF}\simeq 0.572$, and $[g]_{SW}\simeq
2.28$,  $[g]_{SF}\simeq 1.93$, respectively. 
In real-life complex networks with SF structure, however, the  link
conductances are typically {\em weighted} \cite{barrat}, ultimately
leading to better performance for SF networks \cite{Lopez2005}.

\begin{figure}[t]
\vspace*{2.5truecm}
       \includegraphics{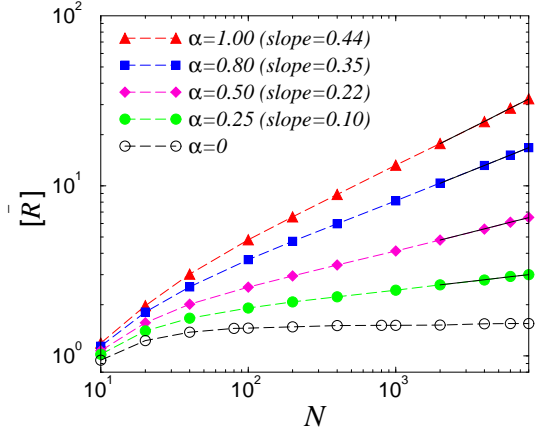}
\vspace*{3.20truecm}
\caption{Average system resistance of SW networks
with distance-dependent conductances as described in the
text, for $p=1.0$ for different $\alpha$ values. The solid line segments
are the measured slopes for $\alpha$$\neq$$0$, with their values shown
in the legends; these values should be compared to $\alpha/2$, the exponent of the leading-order
perturbative results [Eq.~(\ref{R_alpha_N})].}
\label{fig4}
\end{figure}

{\it Effective Resistance of SW Networks with Distance-Dependent Conductances.---}%
Now we consider the more general case where the conductances of the
random (possibly long-range) links decay with the underlying spatial
distance between the nodes they connect in a power-law fashion:  
$J_{ij}=1/|i-j|^{\alpha}$ with probability $p/N$ and $J_{ij}$$=$$0$ with
probability $(1-p/N)$ in Eq.~(\ref{A_sw_ntwk}).
Keeping the density of the random links, $p$, fixed (at a non-zero
value) and taking the large system-size limit, corresponds to the limit
of fixed density of weak links at large scales. Then one can argue
that, to leading order, mean-field scaling holds \cite{KHK05,KHK_SPIE_2005,ee,KHK_PREP,mft}. 
Focusing on the $0$$\leq$$\alpha$$\leq$$1$ regime, we find that
the average link strength, decaying as $[J_{ij}]=(p/N)|i-j|^{-\alpha}$,
gives rise to a system-size dependent effective mass $\Sigma_{N}\simeq (N/2)^{-\alpha}p/(1-\alpha)$
and consequently 
\begin{equation}
[\bar{R}]\simeq \Sigma_{N}^{-1/2}\sim N^{\alpha/2} \;.
\label{R_alpha_N}
\end{equation}
Thus, the average system resistance diverges with the system size
for an arbitrarily small but nonzero value of $\alpha$.
Figure~\ref{fig4} supports this picture, but also indicates that
corrections beyond the mean-field approximation are important
and noticeable for the range of system-sizes that were accessible via
numerical methods.
In fact, an analysis of the naive perturbative
approach \cite{KHK05} reveals that although higher-order corrections are becoming
progressively smaller as $N$ increases, their prefactor is singular
for certain values of $\alpha$ \cite{KHK_PREP}. Given these
subtleties, the deviations [Figure~\ref{fig4}] from the predicted asymptotic scaling
Eq.~(\ref{R_alpha_N}) are reasonable.

For the interested reader, familiar with the diagrammatics of \cite{KHK04},
we present a brief analysis of the higher order corrections.
The mean-field gives a self-energy of order $N^{-\alpha}$.  There are some
corrections for finite $N$ to Eq.~(\ref{R_l}), but
the most important higher-order ones (higher
order in powers of $N^{-1}$ compared to the mean-field)
are corrections to the self-energy.
The leading order correction
to the mean-field involves diagrams in which a single link appears twice;
these diagrams involve summing over the length of the link and are mulitplied
by the strength of the link {\it squared}: $|i-j|^{-2\alpha}$.  For
$\alpha$$<$$1/2$, this sum diverges for large $N$ and gives rise to corrections
to the mean-field $\Sigma$, so that $\Sigma_0=p(N/2)^{-\alpha}/(1-\alpha)$ is
the mean-field value and the leading correction is
$\Sigma=
p(N/2)^{-\alpha}/(1-\alpha)-p
[(N/2)^{-2\alpha}/(1-2\alpha)]/\sqrt{\Sigma_0}+\ldots$. 
Using this correction to $\Sigma$, the corrections to the resistance are
of order $N^0$, and thus for $\alpha$$<$$1/2$ may be significant compared
to the value in Eq.~(\ref{R_alpha_N}).  The coefficient of
these corrections becomes singular at $\alpha$$=$$1/2$ and for $\alpha$$\geq$$1/2$,
the self-energy
$\Sigma$ becomes non-local and acquires
a momentum dependence, which may be shown
to change the form of the higher-order corrections.

Also note that this behavior is very different from that of the case where the
strength (conductance) of the random links is uniform, but the probability of
connecting two nodes, separated by a distance $l$, decays as
$l^{-\alpha}$ \cite{KHK05,KHK_SPIE_2005,JB00}. 
There exists a {\em finite} region, $0$$<$$\alpha$$<$$2$, where the
propagator, hence in the context of this paper, the average system
resistance, remains finite in the limit of $N$$\to$$\infty$. In the present
case, where the link-length distribution is uniform, but the link
strength decays as $l^{-\alpha}$, the average system
resistance is finite {\em only} for $\alpha$$=$$0$.

This contrasting behavior between the two different ``$l^{-\alpha}$''
implementations of the random links (strength vs probability) is in
accord with recent studies on phase transition {\em on} SW
networks. Interacting systems often exhibit mean-field-like 
phase transitions \cite{mft,SCALETT,BARRAT,GITTERMAN,xy_sw,ising_sw,HERRERO02,MARK}, 
even for an arbitrarily small but nonzero density
of random links added to a one-dimensional regular graph.
However, in the case of the strength of the random links decaying in the above
$l^{-\alpha}$ fashion, for the Ising model on SW networks, it was shown \cite{JONG2003}
that no phase transition occurs at any finite
temperature for any nonzero $\alpha$.

{\it Summary and Outlook.---}%
We obtained the scaling behavior of the effective resistance
of SW networks. For uniform link conductances, we found that for an
arbitrary small density of random links, the average system resistance
is finite, and the two-point resistance, as function of the distance
between the nodes, saturates exponentially fast to the same finite value. 
When the link conductance decays with the distance as $l^{-\alpha}$,
the average network resistance diverges with the number of nodes as
$N^{\alpha/2}$.

Ultimately, one is interested not only in the global
transport or flow characteristics of the network, but also in their
effect on the local ``components'', capacity limitations, and possible global network
failures. 
In the context of resistor networks, the question of
voltage landscapes in the network, or more specifically, the voltage-drop distribution
across the links, can be addressed. Such a study can reveal 
the most vulnerable links/connections to be ``blown'' when
increasing the overall load in the network. In
particular, studying the properties of the extreme (largest) voltage-drops
across the links in the network carries information on the weakest
links of the network, and in turn, provides solutions from a
system-design viewpoint. Fuse networks have been intensively studied
on random percolating lattices with various applications to breakdown
processes in condensed matter and materials systems, ranging from
brittle fracture to dielectric breakdown \cite{Hansen91,Herrmann_rev,fuse1,DUXBURY95,fuse2}. 
Future work will address these questions from a general complex network
vulnerability viewpoint \cite{BarabERROR,Havlin00,Havlin01}.

{\it Acknowledgments.---}%
GK and BK were supported by NSF Grant No. DMR-0426488 and the Research
Corporation, MJB and DA by the Australian Research
Council (ARC), MBH by US DOE W-7405-ENG-36, and KEB by NSF Grant No. DMR-0427538 and the Alfred P.
Sloan Foundation.


\begin{thebibliography}{00}


\bibitem{Kirkpatrick71}
S. Kirkpatrick,
Phys. Rev. Lett.\ {\bf 27}, 1722 (1971).

\bibitem{Kirkpatrick73}
S. Kirkpatrick,
Rev.\ Mod.\ Phys.\ {\bf 45}, 574 (1973).

\bibitem{DERRIDA82}
B. Derrida and J. Vannimenus,
J.\ Phys.\ A {\bf 15}, L557-L564 (1982).

\bibitem{Harris86}
A. B. Harris and T. C. Lubensky,
Phys. Rev. B {\bf 35}, 6964 (1987).

\bibitem{Hansen91}
A. Hansen and E.L. Hinrichsen, 
Phys. Rev. B {\bf 43}, 665 (1991).

\bibitem{Herrmann_rev}
{\it Statistical Models for the Fracture of Disordered Media}
edited by H.J. Herrmann and S. Roux (Elsevier, Amsterdam,, 1990).

\bibitem{fuse1}
L. de Arcangelis, S. Redner, and H.J. Herrmann,
J. de Physique {\bf 46}, L585 (1985).

\bibitem{DUXBURY95}
P.M. Duxbury, P.D. Beale, and C. Moukarzel,
Phys. Rev. B {\bf 51}, 3476--3488 (1995).

\bibitem{fuse2}
G.G. Batrouni, A. Hansen and G.H. Ristow,  
J. Phys. A 27, 1363 (1994).


\bibitem{BarabREV}
R. Albert and A.-L. Barab\'asi,
Rev. Mod. Phys. {\bf 74}, 47 (2002).

\bibitem{MendesREV}
S.N. Dorogovtsev and J.F.F. Mendes,
Adv. in Phys. {\bf 51}, 1079 (2002).

\bibitem{NEWMAN_SIAM}
M.E.J. Newman,
SIAM Review {\bf 45}, 167 (2003).


\bibitem{Strogatz_review}
S.H. Strogatz, Nature {\bf 410}, 268 (2001).

\bibitem{BARAHONA02}
M. Barahona and L.M. Pecora,
Phys. Rev. Lett. {\bf 89}, 054101 (2002).

\bibitem{LAI03}
T. Nishikawa, A.E. Motter, Y.-C. Lai, and F.C. Hoppensteadt,
Phys. Rev. Lett. {\bf 91}, 014101 (2003).

\bibitem{MOTTER05}
A.E. Motter, C. Zhou, and J. Kurths,
Phys. Rev. E. {\bf 71}, 016116 (2005).

\bibitem{GRIN05}
G. Grinstein and R. Linsker,
Proc. Natl. Acad. Sci. USA {\bf 102}, 9948 (2005).

\bibitem{KORNISS03}
G. Korniss, M.A. Novotny, H. Guclu, and Z. Toroczkai, P.A.
Rikvold, Science {\bf 299}, 677 (2003).

\bibitem{BARA03}
M. Argollo de Menezes and A.-L. Barab\'asi,
Phys. Rev. Lett. {\bf 92} 028701 (2004).


\bibitem{TORO04}
Z. Toroczkai and K. Bassler, 
Nature {\bf 428}, 716 (2004).

\bibitem{barrat}
A. Barrat, M. Barthelemy, R. Pastor-Satorras, and A. Vespignani,
Proc. Natl. Acad. Sci. USA {\bf 101}, 3747 (2004).

\bibitem{LAI05}
K. Park, Y.-C. Lai, L. Zhao, and N. Ye,
Phys. Rev. E {\bf 71}, 065105(R) (2005).


\bibitem{Berryman04}
M.J.~Berryman, A. Allison, and D. Abbott,
in {\it Noise in Communication}, edited by L.B. White, 
Proceedings of SPIE Vol. 5473 (SPIE, Bellingham, WA, 2004) pp.122--130.

\bibitem{Ashton05} 
D.J. Ashton, T.C. Jarrett, and N.F. Johnson,
Phys. Rev. Lett. {\bf 94}, 058701 (2005).

\bibitem{Aitchison}
R. E. Aitchison,
Am. J. Phys. {\bf 32}, 566 (1964).


\bibitem{NEWMAN01}
M.E.J. Newman,
Phys. Rev. E {\bf 64}, 016132 (2001).

\bibitem{NEWMAN04}
M.E.J. Newman and M. Girvan,
Phys. Rev. E {\bf 69}, 026113 (2004).

\bibitem{HUBER04}
F. Wu and B.A. Huberman, 
Eur. Phys. J. B. {\bf 38}, 331 (2004).

\bibitem{NEWMAN05} 
M.E.J. Newman, 
Social Networks {\bf 27}, 39 (2005).

\bibitem{Lopez2005}
E. L\'opez, S.V. Buldyrev, S. Havlin, and H.E. Stanley,
Phys. Rev. Lett. {\bf 94}, 248701 (2005).

\bibitem{Rieger05}
D.-S. Lee, H. Rieger,
arXiv:cond-mat/0503008 (2005);

\bibitem{Barab_sci}
A.-L. Barab\'asi and R. Albert, 
Science {\bf 286}, 509 (1999).



\bibitem{WATTS98}
D.J. Watts and S.H. Strogatz, Nature {\bf 393}, 440 (1998).

\bibitem{WATTS99}
D.J. Watts,
Am.\ J.\ Soc.\ {\bf 105}, 493 (1999).

\bibitem{NEWMAN}
M.E.J. Newman, J. Stat. Phys. {\bf 101}, 819 (2000).





\bibitem{KHK04} 
B. Kozma, M. B. Hastings, and G. Korniss, 
Phys. Rev. Lett.  {\bf 92}, 108701 (2004).

\bibitem{KHK05} 
B. Kozma, M. B. Hastings, and G. Korniss, 
Phys. Rev. Lett.  {\bf 95}, 018701 (2005).

\bibitem{KHK_SPIE_2005}
B. Kozma, M.B. Hastings, and G. Korniss,
in {\it Noise in Complex Systems and Stochastic Dynamics III},
edited by L.B. Kish, K. Lindenberg, Z. Gingl,
Proceedings of SPIE Vol. 5845 (SPIE, Bellingham, WA, 2005) pp.130--138.

\bibitem{ee} 
M. B. Hastings, Eur. Phys. J. B {\bf 42}, 297 (2004).

\bibitem{EW} 
S.F. Edwards and D.R. Wilkinson, 
Proc. R. Soc. London, Ser A {\bf 381}, 17 (1982).


\bibitem{TORO_VIRTUAL}
Z. Toroczkai, G. Korniss, M. A. Novotny, and H. Guclu,
in {\it Computational Complexity and Statistical Physics}, 
edited by A. Percus, G. Istrate, and C. Moore, 
Santa Fe Institute Studies in the Sciences of Complexity Series  
(Oxford University Press, 2005, in press); arXiv:cond-mat/0304617.

\bibitem{Dong2005}
A. Nagurney, J. Cruz, J. Dong, and D. Zhang,
Eur. J. Oper. Res. {\bf 26}, 120 (2005).


\bibitem{KIRK2004}
S. Kirkpatrick, 
Science {\bf 299}, 668 (2003).


\bibitem{WU2004}
F.Y. Wu,
J.\ Phys.\ A {\bf 37}, 6653 (2004).

\bibitem{Cserti2000}
J. Cserti,
Am.\ J.\ Phys.\ {\bf 68}, 896 (2000).


\bibitem{Doyle}
P.G. Doyle and J.L. Snell, 
{\it Random Walks and Electric Networks},
Carus Mathematical Monograph Series Vol. 22 (The Mathematical
Association of America, Washington, DC, 1984), pp.\ 83--149;
arXive: math.PR/0001057.

\bibitem{Lovasz}
L. Lov\'asz,
{\it Random Walks on Graphs: A Survey in Combinatorics}, Paul Erd\H{o}s is
Eighty Vol. 2, edited by D. Mikl\'os, V.T. S\'os, and T. Sz\H{o}nyi 
(J\'anos Bolyai Mathematical Society, Budapest, 1996), pp.\ 353-398;
http://research.microsoft.com/users/lovasz/erdos.ps.

\bibitem{Redner}
S. Redner,
{\it A Guide to First-Passage Processes} (Cambridge University Press,
Cambridge, UK, 2001).


\bibitem{BARA95}
A.-L. Barab{\'a}si and H.E. Stanley, {\it Fractal Concepts in
Surface Growth} (Cambridge University Press, Cambridge, 1995).

%
%

\bibitem{TORO04a}
M. Anghel, Z. Toroczkai, K.E. Bassler, and G. Korniss,
Phys.\ Rev.\ Lett.\ {\bf 92}, 058701 (2004).

\bibitem{ER}
P. Erd\H{o}s and A. R\'enyi,
Publ.\ Math.\ Inst.\ Hung.\ Acad.\ Sci.\ {\bf 5}, 17 (1960).


\bibitem{NEWMAN_WATTS}
M.E.J. Newman and D.J. Watts, Phys. Lett. A {\bf 263}, 341 (1999).

\bibitem{MONA}
R. Monasson, Eur. Phys. J. B {\bf 12}, 555 (1999).





\bibitem{numrec}
W.H. Press, S.A. Teukolsky, W.T. Vetterling, B.P. Flannery, {\it
Numerical Recipes in C}, 2nd ed. (Cambridge Univ. Press,
Cambridge, 1995), Secs. 11.2 and 11.3.


\bibitem{KHK_PREP}
B. Kozma, M.B. Hastings, and G. Korniss, in preparation.


\bibitem{BA_m}
For the BA scale-free model \cite{Barab_sci} (growth and preferential
attachment), each new node is connected to the network with $m$ links,
resulting in an average degree of $2m$ in the large-$N$ limit.


\bibitem{mft} 
M. B. Hastings, Phys. Rev. Lett. {\bf 91}, 098701 (2003).


\bibitem{JB00} 
S. Jespersen and A. Blumen, 
Phys. Rev. E {\bf 62}, 6270 (2000).


\bibitem{SCALETT}
R. T. Scalettar,
{\it Physica A} {\bf 170}, 282 (1991).

\bibitem{BARRAT}
A. Barrat and M. Weigt, Eur. Phys. J. B {\bf 13}, 547 (2000).

\bibitem{GITTERMAN}
M. Gitterman, J. Phys. A {\bf 33}, 8373 (2000).

\bibitem{xy_sw}
B.J. Kim, H. Hong, P. Holme, G.S. Jeon, P. Minnhagen, and M.Y. Choi, 
Phys. Rev. E {\bf 64}, 056135 (2001).


\bibitem{ising_sw}
H. Hong, B.J. Kim, and M.Y. Choi, Phys. Rev. E {\bf 66}, 018101
(2002).

\bibitem{HERRERO02}
C. P. Herrero,
Phys.\ Rev.\ E {\bf 65}, (2002) 066110.

\bibitem{MARK}
M.A. Novotny and S.M. Wheeler,
Braz. J. Phys. {\bf 34} 395 (2004); 

\bibitem{JONG2003}
D. Jeong, H. Hong, B.J. Kim, and M.Y. Choi,
Phys. Rev. E {\bf 68}, 027101 (2003).

\bibitem{BarabERROR}
R. Albert. H. Jeong, and A.-L. Barab\'asi,
Nature {\bf 406}, 378 (2000).

\bibitem{Havlin00}
R. Cohen, K. Erez, D. ben-Avraham, S. Havlin,
Phys. Rev. Lett. {\bf 85}, 4626 (2000).

\bibitem{Havlin01}
R. Cohen, K. Erez, D. ben-Avraham, S. Havlin,
Phys. Rev. Lett. {\bf 86}, 3682 (2001).



%
%
%
%
%
%
%
%

\end{thebibliography}
\end{document}